\documentclass[prd,preprint,superscriptaddress,nofootinbib,longbibliography,aps]{revtex4-1}
\usepackage[utf8]{inputenc}
\usepackage{graphicx} 
\usepackage{calligra,amsmath,amsfonts,bbm,mathrsfs,amssymb}
\usepackage{graphicx}
\usepackage{caption}
\usepackage{enumerate}
\usepackage{hyperref}
\usepackage{braket}
\usepackage{physics}
\usepackage[dvipsnames]{xcolor}
\usepackage{colortbl}
\usepackage{tabularx}

\begin{document}

\title{Non-invertible symmetry as an axion-less solution to the strong CP problem }

\author{Qiuyue Liang}\email{qiuyue.liang@ipmu.jp}
\affiliation{Kavli IPMU (WPI), UTIAS, University of Tokyo, Kashiwa, 277-8583, Japan}

\author{Tsutomu T. Yanagida}\email{tsutomu.tyanagida@sjtu.edu.cn}
\affiliation{Kavli IPMU (WPI), UTIAS, University of Tokyo, Kashiwa, 277-8583, Japan}
\affiliation{Tsung-Dao Lee Institute \& School of Physics and Astronomy, Shanghai Jiao Tong University, China}

\begin{abstract} 
We use a non-invertible symmetry to construct a three-zero texture for the down-type quark mass matrix, which can resolve the strong CP problem without invoking the axion, in four-dimensional spacetime with three quark families in QCD. We assume CP invariance at the fundamental high-energy scale.  
 \end{abstract}

\maketitle

\section{Introduction} 

The strong charge-parity (CP) problem in QCD is the discrepancy between the theoretically predicted $\mathcal{O}(1)$ CP-violating vacuum angle ${\bar \theta}$ and the experimental upper bound of $ {\bar \theta}<\mathcal{O}(10^{-10})$ from electric dipole moment measurements~\cite{Abel:2020pzs}. 
The CP invariance is the most straightforward solution to the strong CP problem. However, CP invariance must be spontaneously broken at some intermediate scale to generate complex mass matrices for quarks. The diagonalization of these mass matrices induces a shift in the physical vacuum angle ${\bar\theta}$ away from zero unless the determinant of mass matrices for the up- and down- type quark, $\det[M_u M_d]$, is real. Thus, the strong CP problem arises in QCD even if CP is an exact symmetry at the fundamental level.  
 
Recently, several models have been proposed to address the problem by enforcing $\det[M_u M_d]$ to be real through modular symmetry considerations \cite{Feruglio:2023uof,Feruglio:2024ytl,Petcov:2024vph}, where supersymmetry (SUSY) plays an important role. 
 However, these models assume two independent Higgs doublets, and it is unclear whether they should always share the same phase in order to render the determinants of the quark mass matrices real. Moreover, the dynamical SUSY breaking can also introduce CP violation phases at low energy.
 Another approach \cite{Liang:2024wbb} uses a six-dimensional spacetime with $\mathbf{T}^2/\mathbb{Z}_3$ orbifold compactification, but it is unclear whether such compactification successfully avoids generating CP violation at low energies. 

In this short paper we propose a new solution to the strong CP problem in QCD within the standard model framework at the low energy in four dimensional spacetime. We show that a non-invertible symmetry plays a crucial role to satisfy the required condition on mass matrices, that is, arg$({\text{det}[M_u M_d]})$=0.

We begin our discussion of the importance of non-invertible symmetry by considering an interesting mass matrix originally proposed by Weinberg \cite{Weinberg:1977hb},  
\begin{equation} 
\label{eq,weinberg}
      \bar q_L H d_R : \left(\begin{array}{cc}
 0 & \delta  \\
\delta   & M 
\end{array}\right)  \ .
\end{equation}  
This $2\times 2$ mass matrix for the down and strange quarks can explain the Cabibbo mixing angle $\theta_{C}$ \cite{PhysRevLett.10.531} in terms of the mass ratio as $\theta_{C}\simeq \sqrt{m_d/m_s}$ (see also \cite{GATTO1968128}). This is a very successful formula, however, normal symmetry cannot explain the vanishing $(1,1)$ element if the other three matrix elements are non-zero, which is the nature of the group-like symmetry.

However, recent developments of non-group-like symmetries in 4D spacetime have shed light on this problem~ \cite{Tachikawa:2017gyf,Bhardwaj:2022yxj,Bhardwaj:2022lsg,PhysRevLett.128.111601,PhysRevD.105.125016,Kaidi:2022uux, Choi:2022fgx, Hsin:2024aqb,Kaidi:2024wio,Cao:2025qhg} (also see a recent review in \cite{Shao:2023gho}). Non-invertible symmetry is one such symmetry that does not have group structure, which makes them good candidate to explain zero textures in the Yukawa matrix \cite{Kobayashi:2025znw, Kobayashi:2024cvp}. It is a well-motivated concept that has attracted significant interest in recent years with various applications in particle physics \cite{Choi:2022jqy,Cordova:2022qtz,Cordova:2022ieu,Cordova:2022fhg,Cordova:2024ypu,Choi:2023pdp,Kobayashi:2024yqq}. This brings new hope to solve strong CP problem in 4D.

In Sec.\ref{sec:2by2matrix}, we first use a $2 \times 2$ matrix in the Weinberg model as a simple example to illustrate the application of non-invertible symmetry. Then, in Sec.\ref{sec:QCD}, we generalize the mass matrix to $3\times 3$ in the QCD with three families, providing an example of a phenomenological model that can solve the strong CP problem. We conclude in Sec.~\ref{sec:conclusions} with discussions about how the model can be generalized in future studies.

\section{Weinberg model}
\label{sec:2by2matrix} 
We start by discussing the Weinberg model and explain why ordinary symmetry cannot give the desired matrix structure. 

Suppose the left-handed doublet carries charge $(a, b)$ under a certain group-like symmetry, the right-handed singlet has charge $(a', b')$, and the Higgs is neutral. Then, the charge of the Yukawa mass matrix takes the form:
\begin{equation}
      [\bar q_L H d_R] : \left(\begin{array}{cc}
 a+a' & a+b'    \\
a'+b  & b+b'
\end{array}\right) \ ,
\end{equation}
where each element denotes the charge of the couplings. To reproduce the non-vanishing $(1,2), (2,1)$ and $(2,2)$ elements in Eq.~\eqref{eq,weinberg}, the charges must satisfy $a+b' = a'+b = b+b' = 0 $ to allow the corresponding Yukawa couplings. This automatically leads to $a + a' = 0$, which indicates that the $(1,1)$ element has to be allowed by the group-like symmetry. Therefore, the vanishing of the $(1,1)$ element cannot be achieved using an ordinary symmetry in four dimensions.
 
However, symmetries that lack group structures are described by fusion algebra, which allows for different couplings. One way to realize non-invertible symmetries is by gauging the outer automorphism of a non-Abelian group, whose coset\footnote{It is a double coset when considering gauging a non-Abelian group, see \cite{Cao:2025qnc} for a detailed discussion. Inevitably, there will also be mixing between zero-form symmetries and higher-form symmetries in this high category math structure. However, throughout this paper, we focus on the charges that local fields can carry; therefore, only the zero-form symmetries are relevant. } generally exhibits non-invertible symmetry \cite{Bhardwaj:2022lsg,Hsin:2024aqb,Cao:2025qnc}. Non-invertible symmetries are described by fusion rules, enabling different types of couplings \cite{Kaidi:2024wio}. In this paper, we are going to study gauging $\mathbb{Z}_2$ of $\mathbb{Z}_M \rtimes \mathbb{Z}_2 $. This gauging operation identifies charge $k$ with $-k$, and we denote the charge conjugate class as $[g^k]$,
\begin{eqnarray}
    [g^k] = \{h g^k h^{-1}| h \in \mathbb{Z}_2 \}\ ,~ k = 0, 1,\cdots \left\lfloor \frac{M}{2} \right\rfloor \ ,
\end{eqnarray}
where$\left\lfloor x \right\rfloor$ denotes the floor function, which gives the greatest integer that is less than or equal to $x$. 
They follow a fusion rule that is different from the group operation,
\begin{eqnarray}
\label{eq,fusionrule}
  [g^{k_1}][ g^{k_2}]\cdots [g^{k_n}  ] = [g^{k_1 + k_2 +\cdots k_n} ] + [g^{M-k_1 + k_2 +\cdots k_n} ] + [g^{k_1 +M- k_2 +\cdots k_n} ]   +\cdots + [g^{k_1 + k_2 +\cdots M- k_n} ] \ .
\end{eqnarray}
In other words, the $n$-point couplings are allowed in $\mathbb{Z}_M$ if $\sum_i^n k_i = 0 ~(\text{mod} ~ M)$, but are generalized to the condition $\sum_{i}^n \pm k_i = 0~ (\text{mod} ~ M)$ in the non-invertible symmetry case \cite{Kobayashi:2025znw}. Namely, non-invertible symmetries allow more coupling terms than the original group. Consider $M = 3$: we see that the three-point coupling $[g^0][g^1][g^1]$ is now allowed, since $0+1-1=0$, as opposed to the case in ordinary $\mathbb{Z}_3$. This will be the key observation for constructing the matrix in Eq.\eqref{eq,weinberg}. We denote the symmetry with fusion rule Eq.\eqref{eq,fusionrule} of this coset as $\widetilde{\mathbb{Z}}_M$ in what follows.

Now, we assume a non-invertible symmetry $\widetilde{\mathbb{Z}}_3$, which features two classes of charges: $[g^0], [g^1]$.  
We assign the charges of the quarks and the Higgs as follows, 
\begin{equation}
    \bar q_L = ( [g^0] ,[g^1])\ , H = [g^1]\ ,  d_R = ([g^0],[g^1])^T \ , u_R = ([g^1],[g^0])^T\ ,
\end{equation}
where $T$ denotes the transverse, and 
\begin{eqnarray}
 \bar q_L =  \left(\begin{array}{cc}
u & c    \\
d &s
\end{array}\right)_L\ , \   d_R = \left(\begin{array}{c}
d      \\
s  
\end{array}\right)_R  \ ,\  u_R = \left(\begin{array}{c}
u      \\
c  
\end{array}\right)_R   \ .
\end{eqnarray}
 The mass matrix naturally turns to be,
\begin{equation}
\label{eq,weinbergmatrix}
  \bar q_L H^\dagger d_R :    \left(\begin{array}{cc}
0 & \checkmark   \\
\checkmark & \checkmark     
\end{array}\right)\ , \quad    \bar q_L H u_R  :   \left(\begin{array}{cc}
\checkmark & 0    \\
\checkmark & \checkmark 
\end{array}\right)\  ,
\end{equation}
with the down quark sector fully recovers the desired Weinberg matrix in Eq.~\eqref{eq,weinberg}, where $\checkmark$ denotes the allowed couplings. The up quark sector has a non-vanishing $(2,1)$ element from the coupling between the left-handed charm and right-handed up quarks, but when applying the left-handed unitary matrices rotation, this element becomes negligible due to the large mass hierarchy, $m_u / m_c \ll 1$. As a result, we recover the Weinberg mass matrix, with the Cabibbo angle approximately given by $\theta_C \simeq \sqrt{m_d / m_s}$. Note that this relation cannot be achieved using an ordinary symmetry. The non-invertible symmetry $\widetilde{\mathbb{Z}}_3$ provides a simple yet illustrative example of the power of applying non-invertible symmetries in particle physics model building.  

It should be noted that the quantum corrections can induce the $(1,1)$ element in the down-type quark mass matrix in Eq.\eqref{eq,weinbergmatrix}\footnote{This is an example where the 't Hooft naturalness condition \cite{tHooft:1979rat} is not applicable.}, but it is negligibly small.

\section{QCD with three families}
\label{sec:QCD} 
With the knowledge learned in the last section, we now generalize to the QCD with three families and discuss the quark matrices. As discussed in the paper \cite{Liang:2024wbb}, what we need is a charge assignment so that the mass matrix of the up quark sector is almost diagonal, and that of the down quark sector has three-zero texture so as to have a real determinant while presenting the CP violation phase \cite{Tanimoto:2016rqy} in the Cabibbo-Kobayashi-Maskawa(CKM) matrix. The three-zero texture cannot be achieved by ordinary symmetries in 4D (\cite{Liang:2024wbb} ) following similar argument in Sec.~\ref{sec:2by2matrix}, but we will show that a non-invertible symmetry with non-trivial fusion rules can accomplish this. 

We consider a two Higgs doublet model, where both of the light Higgs doublets, $H_1$ and $H_2$, gain complex vevs in general if the CP is broken at the intermediate energy scale, $\Lambda\sim 10^{12}$ GeV, for instance. However, we can make the vev of $H_1$ real by the $U(1)_Y$ gauge rotation, but the vev of $H_2$ remains complex, which contributes to the CP violation phase observed in the kaon decay. The dynamics of the spontaneous breaking of the CP invariance are discussed in Appendix~\ref{sec:appendix2}.

We consider the $\widetilde{\mathbb{Z}}_5 \times \widetilde{\mathbb{Z}}_3\times \widetilde{\mathbb{Z}}_3 $ symmetry, where generator in $\widetilde{\mathbb{Z}}_5$ and $\widetilde{\mathbb{Z}}_3$ have charges 
\begin{equation}
    [g_5^k]: \  k = 0,1,2\ ,~ [g_3^k]: \  k = 0,1 \ .
\end{equation}
The allowed three-point couplings of $\widetilde{\mathbb{Z}}_5$ are $([g_5^0][g_5^0][g_5^0]), ([g_5^0][g_5^1][g_5^1]), ([g_5^0][g_5^2][g_5^2])$, $([g_5^1][g_5^1][g_5^2])$ and $([g_5^1][g_5^2][g_5^2])$.  
We assign the charge as following: 
\begin{equation}
\label{eq,charge1}
    [\bar q_L] = ( [g_5^0][g_3^1][g_3^0], [g_5^1] [g_3^1] [g_3^1] , [g_5^2][g_3^1][g_3^0])   \ , [H_1]  = [g_5^1][g_3^1][g_3^1] \ , [H_2]  = [g_5^0][g_3^0][g_3^0] 
\end{equation}
\begin{equation}
\label{eq,charge2}
 [ u_R]=([g_5^1][g_3^0][g_3^1], [g_5^0][g_3^0][g_3^0], [g_5^2][g_3^0][g_3^1])^T \ , 
 [d_R ] = ( [g_5^0][g_3^0][g_3^1], [g_5^1][g_3^1][g_3^1], [g_5^2][g_3^0][g_3^1])^T \ ,
\end{equation}
where the first bracket denotes the charge for $\widetilde{\mathbb{Z}}_5$, and the other two is for $\widetilde{\mathbb{Z}}_3$.

Therefore, we have up quark Yukawa matrix looks like 
\begin{equation}
\label{eq,upquarkmatrix}
        \bar u_L H_1 u_R : \left(\begin{array}{ccc}
\checkmark & 0 & 0 \\
0 & \checkmark & \checkmark \\
\checkmark & 0 & \checkmark
\end{array}\right)\ , \quad   \bar u_L H_2  u_R :  \left(\begin{array}{ccc}
0 & 0 & 0 \\
0 & 0 & 0 \\
0 & 0 & 0
\end{array}\right)\ . 
\end{equation}
For down quark sector, we have 
\begin{equation}
\label{eq,downquarkmatrix}
   \bar d_L H_1^\dagger d_R : \left(\begin{array}{ccc}
0 & \checkmark & 0 \\
\checkmark & 0 & \checkmark \\
0 & \checkmark & \checkmark
\end{array}\right)\ , \quad   \bar d_L H_2^\dagger  d_R :  \left(\begin{array}{ccc}
0 & 0 & 0 \\
0 & \star & 0 \\
0 & 0 & 0
\end{array}\right)\ .
\end{equation} 
Here, $\checkmark$ denotes allowed real coupling coefficients, and $\star$ denotes complex ones.   There are ten real degrees of freedom and one complex ones. Notice that the three-zero texture can be achieved by other non-invertible symmetries, as discussed in \cite{Kobayashi:2025znw}; however, they failed to explain the localized complex phase which is crucial to solve the strong CP problem. 

We emphasize three key points. First, the determinants of both the up- and down-type quark mass matrices are real, thereby solving the strong CP problem in QCD. Second, as discussed in Sec.\ref{sec:2by2matrix}, the $(3,1)$ element in the up-type quark mass matrix can be safely neglected when analyzing CKM phenomenology because of the mass hierarchy $m_u/m_t \ll 1$. As for the $(2,3)$ element we can  prove the corresponding mixing is sufficiently suppressed if it is smaller than $\sim 0.01\times m_t$ (see \cite{Tanimoto:2016rqy}).  As a result, the up-type quark mass matrix can effectively be treated as a real diagonal matrix if the above condition is satisfied. Third, the down-type quark mass matrix is identical to that used in \cite{Tanimoto:2016rqy,Liang:2024wbb}, and we have confirmed that it can successfully reproduce the observed quark masses as well as the full CKM matrix\cite{Tanimoto:2016rqy}, including the CP-violating phase.

We should also take quantum corrections into consideration. At the one-loop level, there is only wave-function renormalization for the quark fields, which does not induce any complex phase in $\det[M_d M_u]$, since the wave-function renormalization factors $Z$ are Hermitian matrices \cite{Hiller:2001qg}. At the two-loop level, vertex corrections to the Yukawa coupling matrices can generate a complex phase in $\det[M_d M_u]$, but the resulting shift in the vacuum angle, $\Delta {\bar \theta} \sim 10^{-15}$, is well below the current experimental bound ${\bar \theta} \leq 10^{-10}$ \cite{Abel:2020pzs}.

\section{Conclusions and discussion}
\label{sec:conclusions} 
In this paper, we construct a three-zero texture for the down type quark mass matrix to address the strong CP problem using a non-invertible symmetry $ \widetilde{\mathbb{Z}}_5 \times \widetilde{\mathbb{Z}}_3 \times \widetilde{\mathbb{Z}}_3  $, obtained by gauging the outer automorphism $\mathbb{Z}_2$ of $\mathbb{Z}_{5,3} \rtimes \mathbb{Z}_2$, respectively. We first explain how to apply the non-invertible symmetry to a $2 \times 2$ matrix in the Weinberg model, and then include the three family case. 
The model can be straightforwardly extended to the lepton sector, where CP violation in neutrino oscillations~\cite{Tanimoto:2024nwm,Tanimoto:2025fnj} could serve as a potential prediction of this non-invertible symmetry. This extension will be addressed in future work.

In the present model, the Standard Model Higgs doublet $H$ is a linear combination of $H_1$ and $H_2$, while the orthogonal combination corresponds to a heavy Higgs, $H'$, with mass $m^\prime \sim \mathcal{O}(10)$ TeV. The physics at the weak scale is not affected by the heavy Higgs and is well described within the Standard Model~\cite{Branco:2011iw}. However, flavour-changing neutral current decays of the $b$ and $s$ quarks may serve as a smoking gun for the present model.

Notice that the mathematical structure of the coset arising from gauging an automorphism of a group $G$ is generally described by a higher-category structure, where mixing between non-invertible symmetries and higher-form symmetries typically occurs. Whether this will lead to interesting phenomenology is left for future discussion. It is also interesting to explore other possible non-invertible symmetries besides $\widetilde{\mathbb{Z}}_{5}\times \widetilde{\mathbb{Z}}_3\times \widetilde{\mathbb{Z}}_3 $, and to study the origin of such non-invertible symmetries.

\section*{Ackowledgments}
We thank Weiguang Cao, Tom Melia, Morimitsu Tanimoto, Hao Y. Zhang, Yi Zhang for helpful discussion. We thank Weiguang Cao for useful comment on the draft. This work is supported by MEXT KAKENHI Grants No.~24H02244 (T.~T.~Y.) and the Natural Science Foundation of China (NSFC) under Grant No.~12175134 as well as by World Premier International Research Center Initiative (WPI Initiative), MEXT, Japan.

\appendix  
\section{Effective action for Higgs}
\label{sec:appendix2}
We discuss a possible UV realization of the two Higgs doublet model discussed in the main context. 

The potential of Higgs takes the form
\begin{eqnarray}
\label{eq,Higgspotential}
     V(H_1,H_2)  &= & - m_1^2 H_1H_1^\dagger- m_2^2 H_2H_2^\dagger  +\lambda_1\left(H_{1}^{\dagger} H_{1}\right)^2  + \lambda_2\left(H_{2}^{\dagger} H_{2}\right)^2   \nonumber \\
&&+ 
\alpha_1\left(H_{1}^{\dagger} H_2 \langle\eta\rangle+H_2^{\dagger} H_{1} \langle\eta^{\dagger}\rangle\right) +\alpha_2\left(H_1^{\dagger} H_{2} \langle\eta\rangle+H_{2}^{\dagger} H_1 \langle\eta^{\dagger}\rangle\right)\nonumber \\
&& +\beta_1 H_1^{\dagger} H_1 H_{2}^{\dagger} H_{2}+\beta_2\left[\left(H_1^{\dagger} H_{2}\right)^2+\left(H_{2}^{\dagger} H_1\right)^2\right] \ . 
\end{eqnarray}
while all coefficients beside the $\langle\eta\rangle$ are real since CP is fundamentally invariant. Here, $m_{1,2} \sim \mathcal{O}(10)$ TeV are the masses of the two Higgs bosons, and $\alpha_{1,2} $ are dimension-one coefficients that generate complex mixing between $H_{1,2}$ when complex scalar field $\eta$ acquires a vev, $\langle\eta\rangle$, as CP is spontaneously broken at $\Lambda \sim 10^{12}$ GeV. We need $\alpha_{1,2} \langle\eta\rangle \sim \mathcal{O}(100) ~\text{TeV}^2$.  

The potential of $\eta$ is,  
\begin{eqnarray}
\label{eq,etapotential}
    V (\eta)=-\mu_1 \eta \eta^{\dagger}-\mu_2\left(\eta^2+\eta^{\dagger 2}\right)+\xi_1\left(\eta \eta^{\dagger}\right)^2+\xi_2\left(\eta^4+\eta^{\dagger 4}\right) \ .
\end{eqnarray}
We have a non-vanishing complex vev for the $\eta$ in the vacuum as shown in \cite{Liang:2024wbb}, which  spontaneously breaks the CP invariance at the scale $\sim 10^{12}$ GeV.
Integrating out the heavy $\eta$ contributes to the complex coefficients in the Higgs potential in Eq.\eqref{eq,Higgspotential} . We can choose the proper $U(1)$ rotation such that $\langle{H_1}\rangle $ is real, while $\langle{H_2}\rangle $ is complex and contributes to the CP violation phase in the CKM matrix. The standard model Higgs doublet $H$ is a linear combination of the $H_1$ and $H_2$ and if we take a normal phase convension where the vev of $H$ is real, both vevs of the $H_{1,2}$ are complex. However, we can see det$[M_d M_u]$ is real. Assigning the proper charge, we therefore obtain the quark mass matrix as discussed in Eq.\eqref{eq,upquarkmatrix} and \eqref{eq,downquarkmatrix}. 

\bibliography{ref}

\begin{thebibliography}{35}%
\makeatletter
\providecommand \@ifxundefined [1]{%
 \@ifx{#1\undefined}
}%
\providecommand \@ifnum [1]{%
 \ifnum #1\expandafter \@firstoftwo
 \else \expandafter \@secondoftwo
 \fi
}%
\providecommand \@ifx [1]{%
 \ifx #1\expandafter \@firstoftwo
 \else \expandafter \@secondoftwo
 \fi
}%
\providecommand \natexlab [1]{#1}%
\providecommand \enquote  [1]{``#1''}%
\providecommand \bibnamefont  [1]{#1}%
\providecommand \bibfnamefont [1]{#1}%
\providecommand \citenamefont [1]{#1}%
\providecommand \href@noop [0]{\@secondoftwo}%
\providecommand \href [0]{\begingroup \@sanitize@url \@href}%
\providecommand \@href[1]{\@@startlink{#1}\@@href}%
\providecommand \@@href[1]{\endgroup#1\@@endlink}%
\providecommand \@sanitize@url [0]{\catcode `\\12\catcode `\$12\catcode
  `\&12\catcode `\#12\catcode `\^12\catcode `\_12\catcode `\%12\relax}%
\providecommand \@@startlink[1]{}%
\providecommand \@@endlink[0]{}%
\providecommand \url  [0]{\begingroup\@sanitize@url \@url }%
\providecommand \@url [1]{\endgroup\@href {#1}{\urlprefix }}%
\providecommand \urlprefix  [0]{URL }%
\providecommand \Eprint [0]{\href }%
\providecommand \doibase [0]{http://dx.doi.org/}%
\providecommand \selectlanguage [0]{\@gobble}%
\providecommand \bibinfo  [0]{\@secondoftwo}%
\providecommand \bibfield  [0]{\@secondoftwo}%
\providecommand \translation [1]{[#1]}%
\providecommand \BibitemOpen [0]{}%
\providecommand \bibitemStop [0]{}%
\providecommand \bibitemNoStop [0]{.\EOS\space}%
\providecommand \EOS [0]{\spacefactor3000\relax}%
\providecommand \BibitemShut  [1]{\csname bibitem#1\endcsname}%
\let\auto@bib@innerbib\@empty
\bibitem [{\citenamefont {Abel}\ \emph {et~al.}(2020)\citenamefont {Abel} \emph
  {et~al.}}]{Abel:2020pzs}%
  \BibitemOpen
  \bibfield  {author} {\bibinfo {author} {\bibfnamefont {C.}~\bibnamefont
  {Abel}} \emph {et~al.},\ }\bibfield  {title} {\enquote {\bibinfo {title}
  {{Measurement of the Permanent Electric Dipole Moment of the Neutron}},}\
  }\href {\doibase 10.1103/PhysRevLett.124.081803} {\bibfield  {journal}
  {\bibinfo  {journal} {Phys. Rev. Lett.}\ }\textbf {\bibinfo {volume} {124}},\
  \bibinfo {pages} {081803} (\bibinfo {year} {2020})},\ \Eprint
  {http://arxiv.org/abs/2001.11966} {arXiv:2001.11966 [hep-ex]} \BibitemShut
  {NoStop}%
\bibitem [{\citenamefont {Feruglio}\ \emph {et~al.}(2023)\citenamefont
  {Feruglio}, \citenamefont {Strumia},\ and\ \citenamefont
  {Titov}}]{Feruglio:2023uof}%
  \BibitemOpen
  \bibfield  {author} {\bibinfo {author} {\bibfnamefont {Ferruccio}\
  \bibnamefont {Feruglio}}, \bibinfo {author} {\bibfnamefont {Alessandro}\
  \bibnamefont {Strumia}}, \ and\ \bibinfo {author} {\bibfnamefont {Arsenii}\
  \bibnamefont {Titov}},\ }\bibfield  {title} {\enquote {\bibinfo {title}
  {{Modular invariance and the QCD angle}},}\ }\href {\doibase
  10.1007/JHEP07(2023)027} {\bibfield  {journal} {\bibinfo  {journal} {JHEP}\
  }\textbf {\bibinfo {volume} {07}},\ \bibinfo {pages} {027} (\bibinfo {year}
  {2023})},\ \Eprint {http://arxiv.org/abs/2305.08908} {arXiv:2305.08908
  [hep-ph]} \BibitemShut {NoStop}%
\bibitem [{\citenamefont {Feruglio}\ \emph {et~al.}(2024)\citenamefont
  {Feruglio}, \citenamefont {Parriciatu}, \citenamefont {Strumia},\ and\
  \citenamefont {Titov}}]{Feruglio:2024ytl}%
  \BibitemOpen
  \bibfield  {author} {\bibinfo {author} {\bibfnamefont {Ferruccio}\
  \bibnamefont {Feruglio}}, \bibinfo {author} {\bibfnamefont {Matteo}\
  \bibnamefont {Parriciatu}}, \bibinfo {author} {\bibfnamefont {Alessandro}\
  \bibnamefont {Strumia}}, \ and\ \bibinfo {author} {\bibfnamefont {Arsenii}\
  \bibnamefont {Titov}},\ }\bibfield  {title} {\enquote {\bibinfo {title}
  {{Solving the strong CP problem without axions}},}\ }\href {\doibase
  10.1007/JHEP08(2024)214} {\bibfield  {journal} {\bibinfo  {journal} {JHEP}\
  }\textbf {\bibinfo {volume} {08}},\ \bibinfo {pages} {214} (\bibinfo {year}
  {2024})},\ \Eprint {http://arxiv.org/abs/2406.01689} {arXiv:2406.01689
  [hep-ph]} \BibitemShut {NoStop}%
\bibitem [{\citenamefont {Petcov}\ and\ \citenamefont
  {Tanimoto}(2024)}]{Petcov:2024vph}%
  \BibitemOpen
  \bibfield  {author} {\bibinfo {author} {\bibfnamefont {S.~T.}\ \bibnamefont
  {Petcov}}\ and\ \bibinfo {author} {\bibfnamefont {M.}~\bibnamefont
  {Tanimoto}},\ }\bibfield  {title} {\enquote {\bibinfo {title} {{$A_4$ modular
  invariance and the strong CP problem}},}\ }\href {\doibase
  10.1140/epjc/s10052-024-13272-w} {\bibfield  {journal} {\bibinfo  {journal}
  {Eur. Phys. J. C}\ }\textbf {\bibinfo {volume} {84}},\ \bibinfo {pages} {914}
  (\bibinfo {year} {2024})},\ \Eprint {http://arxiv.org/abs/2404.00858}
  {arXiv:2404.00858 [hep-ph]} \BibitemShut {NoStop}%
\bibitem [{\citenamefont {Liang}\ \emph {et~al.}(2024)\citenamefont {Liang},
  \citenamefont {Okabe},\ and\ \citenamefont {Yanagida}}]{Liang:2024wbb}%
  \BibitemOpen
  \bibfield  {author} {\bibinfo {author} {\bibfnamefont {Qiuyue}\ \bibnamefont
  {Liang}}, \bibinfo {author} {\bibfnamefont {Risshin}\ \bibnamefont {Okabe}},
  \ and\ \bibinfo {author} {\bibfnamefont {Tsutomu~T.}\ \bibnamefont
  {Yanagida}},\ }\bibfield  {title} {\enquote {\bibinfo {title} {{Three-zero
  texture of quark-mass matrices as a solution to the strong CP problem}},}\
  }\href {\doibase 10.1016/j.physletb.2024.139123} {\bibfield  {journal}
  {\bibinfo  {journal} {Phys. Lett. B}\ }\textbf {\bibinfo {volume} {859}},\
  \bibinfo {pages} {139123} (\bibinfo {year} {2024})},\ \Eprint
  {http://arxiv.org/abs/2408.12146} {arXiv:2408.12146 [hep-ph]} \BibitemShut
  {NoStop}%
\bibitem [{\citenamefont {Weinberg}(1977)}]{Weinberg:1977hb}%
  \BibitemOpen
  \bibfield  {author} {\bibinfo {author} {\bibfnamefont {Steven}\ \bibnamefont
  {Weinberg}},\ }\bibfield  {title} {\enquote {\bibinfo {title} {{The Problem
  of Mass}},}\ }\href {\doibase 10.1111/j.2164-0947.1977.tb02958.x} {\bibfield
  {journal} {\bibinfo  {journal} {Trans. New York Acad. Sci.}\ }\textbf
  {\bibinfo {volume} {38}},\ \bibinfo {pages} {185--201} (\bibinfo {year}
  {1977})}\BibitemShut {NoStop}%
\bibitem [{\citenamefont {Cabibbo}(1963)}]{PhysRevLett.10.531}%
  \BibitemOpen
  \bibfield  {author} {\bibinfo {author} {\bibfnamefont {Nicola}\ \bibnamefont
  {Cabibbo}},\ }\bibfield  {title} {\enquote {\bibinfo {title} {Unitary
  symmetry and leptonic decays},}\ }\href {\doibase 10.1103/PhysRevLett.10.531}
  {\bibfield  {journal} {\bibinfo  {journal} {Phys. Rev. Lett.}\ }\textbf
  {\bibinfo {volume} {10}},\ \bibinfo {pages} {531--533} (\bibinfo {year}
  {1963})}\BibitemShut {NoStop}%
\bibitem [{\citenamefont {Gatto}\ \emph {et~al.}(1968)\citenamefont {Gatto},
  \citenamefont {Sartori},\ and\ \citenamefont {Tonin}}]{GATTO1968128}%
  \BibitemOpen
  \bibfield  {author} {\bibinfo {author} {\bibfnamefont {R.}~\bibnamefont
  {Gatto}}, \bibinfo {author} {\bibfnamefont {G.}~\bibnamefont {Sartori}}, \
  and\ \bibinfo {author} {\bibfnamefont {M.}~\bibnamefont {Tonin}},\ }\bibfield
   {title} {\enquote {\bibinfo {title} {Weak self-masses, cabibbo angle, and
  broken su2 × su2},}\ }\href {\doibase
  https://doi.org/10.1016/0370-2693(68)90150-0} {\bibfield  {journal} {\bibinfo
   {journal} {Physics Letters B}\ }\textbf {\bibinfo {volume} {28}},\ \bibinfo
  {pages} {128--130} (\bibinfo {year} {1968})}\BibitemShut {NoStop}%
\bibitem [{\citenamefont {Tachikawa}(2020)}]{Tachikawa:2017gyf}%
  \BibitemOpen
  \bibfield  {author} {\bibinfo {author} {\bibfnamefont {Yuji}\ \bibnamefont
  {Tachikawa}},\ }\bibfield  {title} {\enquote {\bibinfo {title} {{On gauging
  finite subgroups}},}\ }\href {\doibase 10.21468/SciPostPhys.8.1.015}
  {\bibfield  {journal} {\bibinfo  {journal} {SciPost Phys.}\ }\textbf
  {\bibinfo {volume} {8}},\ \bibinfo {pages} {015} (\bibinfo {year} {2020})},\
  \Eprint {http://arxiv.org/abs/1712.09542} {arXiv:1712.09542 [hep-th]}
  \BibitemShut {NoStop}%
\bibitem [{\citenamefont {Bhardwaj}\ \emph {et~al.}(2023)\citenamefont
  {Bhardwaj}, \citenamefont {Bottini}, \citenamefont {Schafer-Nameki},\ and\
  \citenamefont {Tiwari}}]{Bhardwaj:2022yxj}%
  \BibitemOpen
  \bibfield  {author} {\bibinfo {author} {\bibfnamefont {Lakshya}\ \bibnamefont
  {Bhardwaj}}, \bibinfo {author} {\bibfnamefont {Lea~E.}\ \bibnamefont
  {Bottini}}, \bibinfo {author} {\bibfnamefont {Sakura}\ \bibnamefont
  {Schafer-Nameki}}, \ and\ \bibinfo {author} {\bibfnamefont {Apoorv}\
  \bibnamefont {Tiwari}},\ }\bibfield  {title} {\enquote {\bibinfo {title}
  {{Non-invertible higher-categorical symmetries}},}\ }\href {\doibase
  10.21468/SciPostPhys.14.1.007} {\bibfield  {journal} {\bibinfo  {journal}
  {SciPost Phys.}\ }\textbf {\bibinfo {volume} {14}},\ \bibinfo {pages} {007}
  (\bibinfo {year} {2023})},\ \Eprint {http://arxiv.org/abs/2204.06564}
  {arXiv:2204.06564 [hep-th]} \BibitemShut {NoStop}%
\bibitem [{\citenamefont {Bhardwaj}\ \emph {et~al.}(2022)\citenamefont
  {Bhardwaj}, \citenamefont {Schafer-Nameki},\ and\ \citenamefont
  {Wu}}]{Bhardwaj:2022lsg}%
  \BibitemOpen
  \bibfield  {author} {\bibinfo {author} {\bibfnamefont {Lakshya}\ \bibnamefont
  {Bhardwaj}}, \bibinfo {author} {\bibfnamefont {Sakura}\ \bibnamefont
  {Schafer-Nameki}}, \ and\ \bibinfo {author} {\bibfnamefont {Jingxiang}\
  \bibnamefont {Wu}},\ }\bibfield  {title} {\enquote {\bibinfo {title}
  {{Universal Non-Invertible Symmetries}},}\ }\href {\doibase
  10.1002/prop.202200143} {\bibfield  {journal} {\bibinfo  {journal} {Fortsch.
  Phys.}\ }\textbf {\bibinfo {volume} {70}},\ \bibinfo {pages} {2200143}
  (\bibinfo {year} {2022})},\ \Eprint {http://arxiv.org/abs/2208.05973}
  {arXiv:2208.05973 [hep-th]} \BibitemShut {NoStop}%
\bibitem [{\citenamefont {Kaidi}\ \emph
  {et~al.}(2022{\natexlab{a}})\citenamefont {Kaidi}, \citenamefont {Ohmori},\
  and\ \citenamefont {Zheng}}]{PhysRevLett.128.111601}%
  \BibitemOpen
  \bibfield  {author} {\bibinfo {author} {\bibfnamefont {Justin}\ \bibnamefont
  {Kaidi}}, \bibinfo {author} {\bibfnamefont {Kantaro}\ \bibnamefont {Ohmori}},
  \ and\ \bibinfo {author} {\bibfnamefont {Yunqin}\ \bibnamefont {Zheng}},\
  }\bibfield  {title} {\enquote {\bibinfo {title} {Kramers-wannier-like duality
  defects in $(3+1)d$ gauge theories},}\ }\href {\doibase
  10.1103/PhysRevLett.128.111601} {\bibfield  {journal} {\bibinfo  {journal}
  {Phys. Rev. Lett.}\ }\textbf {\bibinfo {volume} {128}},\ \bibinfo {pages}
  {111601} (\bibinfo {year} {2022}{\natexlab{a}})}\BibitemShut {NoStop}%
\bibitem [{\citenamefont {Choi}\ \emph
  {et~al.}(2022{\natexlab{a}})\citenamefont {Choi}, \citenamefont {C\'ordova},
  \citenamefont {Hsin}, \citenamefont {Lam},\ and\ \citenamefont
  {Shao}}]{PhysRevD.105.125016}%
  \BibitemOpen
  \bibfield  {author} {\bibinfo {author} {\bibfnamefont {Yichul}\ \bibnamefont
  {Choi}}, \bibinfo {author} {\bibfnamefont {Clay}\ \bibnamefont {C\'ordova}},
  \bibinfo {author} {\bibfnamefont {Po-Shen}\ \bibnamefont {Hsin}}, \bibinfo
  {author} {\bibfnamefont {Ho~Tat}\ \bibnamefont {Lam}}, \ and\ \bibinfo
  {author} {\bibfnamefont {Shu-Heng}\ \bibnamefont {Shao}},\ }\bibfield
  {title} {\enquote {\bibinfo {title} {Noninvertible duality defects in $3+1$
  dimensions},}\ }\href {\doibase 10.1103/PhysRevD.105.125016} {\bibfield
  {journal} {\bibinfo  {journal} {Phys. Rev. D}\ }\textbf {\bibinfo {volume}
  {105}},\ \bibinfo {pages} {125016} (\bibinfo {year}
  {2022}{\natexlab{a}})}\BibitemShut {NoStop}%
\bibitem [{\citenamefont {Kaidi}\ \emph
  {et~al.}(2022{\natexlab{b}})\citenamefont {Kaidi}, \citenamefont {Zafrir},\
  and\ \citenamefont {Zheng}}]{Kaidi:2022uux}%
  \BibitemOpen
  \bibfield  {author} {\bibinfo {author} {\bibfnamefont {Justin}\ \bibnamefont
  {Kaidi}}, \bibinfo {author} {\bibfnamefont {Gabi}\ \bibnamefont {Zafrir}}, \
  and\ \bibinfo {author} {\bibfnamefont {Yunqin}\ \bibnamefont {Zheng}},\
  }\bibfield  {title} {\enquote {\bibinfo {title} {{Non-invertible symmetries
  of $ \mathcal{N} $ = 4 SYM and twisted compactification}},}\ }\href {\doibase
  10.1007/JHEP08(2022)053} {\bibfield  {journal} {\bibinfo  {journal} {JHEP}\
  }\textbf {\bibinfo {volume} {08}},\ \bibinfo {pages} {053} (\bibinfo {year}
  {2022}{\natexlab{b}})},\ \Eprint {http://arxiv.org/abs/2205.01104}
  {arXiv:2205.01104 [hep-th]} \BibitemShut {NoStop}%
\bibitem [{\citenamefont {Choi}\ \emph {et~al.}(2023)\citenamefont {Choi},
  \citenamefont {Lam},\ and\ \citenamefont {Shao}}]{Choi:2022fgx}%
  \BibitemOpen
  \bibfield  {author} {\bibinfo {author} {\bibfnamefont {Yichul}\ \bibnamefont
  {Choi}}, \bibinfo {author} {\bibfnamefont {Ho~Tat}\ \bibnamefont {Lam}}, \
  and\ \bibinfo {author} {\bibfnamefont {Shu-Heng}\ \bibnamefont {Shao}},\
  }\bibfield  {title} {\enquote {\bibinfo {title} {{Non-invertible Gauss law
  and axions}},}\ }\href {\doibase 10.1007/JHEP09(2023)067} {\bibfield
  {journal} {\bibinfo  {journal} {JHEP}\ }\textbf {\bibinfo {volume} {09}},\
  \bibinfo {pages} {067} (\bibinfo {year} {2023})},\ \Eprint
  {http://arxiv.org/abs/2212.04499} {arXiv:2212.04499 [hep-th]} \BibitemShut
  {NoStop}%
\bibitem [{\citenamefont {Hsin}\ \emph {et~al.}(2024)\citenamefont {Hsin},
  \citenamefont {Kobayashi},\ and\ \citenamefont {Zhang}}]{Hsin:2024aqb}%
  \BibitemOpen
  \bibfield  {author} {\bibinfo {author} {\bibfnamefont {Po-Shen}\ \bibnamefont
  {Hsin}}, \bibinfo {author} {\bibfnamefont {Ryohei}\ \bibnamefont
  {Kobayashi}}, \ and\ \bibinfo {author} {\bibfnamefont {Carolyn}\ \bibnamefont
  {Zhang}},\ }\bibfield  {title} {\enquote {\bibinfo {title}
  {{Fractionalization of coset non-invertible symmetry and exotic Hall
  conductance}},}\ }\href {\doibase 10.21468/SciPostPhys.17.3.095} {\bibfield
  {journal} {\bibinfo  {journal} {SciPost Phys.}\ }\textbf {\bibinfo {volume}
  {17}},\ \bibinfo {pages} {095} (\bibinfo {year} {2024})},\ \Eprint
  {http://arxiv.org/abs/2405.20401} {arXiv:2405.20401 [cond-mat.str-el]}
  \BibitemShut {NoStop}%
\bibitem [{\citenamefont {Kaidi}\ \emph {et~al.}(2024)\citenamefont {Kaidi},
  \citenamefont {Tachikawa},\ and\ \citenamefont {Zhang}}]{Kaidi:2024wio}%
  \BibitemOpen
  \bibfield  {author} {\bibinfo {author} {\bibfnamefont {Justin}\ \bibnamefont
  {Kaidi}}, \bibinfo {author} {\bibfnamefont {Yuji}\ \bibnamefont {Tachikawa}},
  \ and\ \bibinfo {author} {\bibfnamefont {Hao~Y.}\ \bibnamefont {Zhang}},\
  }\bibfield  {title} {\enquote {\bibinfo {title} {{On a class of selection
  rules without group actions in field theory and string theory}},}\ }\href
  {\doibase 10.21468/SciPostPhys.17.6.169} {\bibfield  {journal} {\bibinfo
  {journal} {SciPost Phys.}\ }\textbf {\bibinfo {volume} {17}},\ \bibinfo
  {pages} {169} (\bibinfo {year} {2024})},\ \Eprint
  {http://arxiv.org/abs/2402.00105} {arXiv:2402.00105 [hep-th]} \BibitemShut
  {NoStop}%
\bibitem [{\citenamefont {Cao}\ \emph {et~al.}(2025{\natexlab{a}})\citenamefont
  {Cao}, \citenamefont {Yamazaki},\ and\ \citenamefont {Li}}]{Cao:2025qhg}%
  \BibitemOpen
  \bibfield  {author} {\bibinfo {author} {\bibfnamefont {Weiguang}\
  \bibnamefont {Cao}}, \bibinfo {author} {\bibfnamefont {Masahito}\
  \bibnamefont {Yamazaki}}, \ and\ \bibinfo {author} {\bibfnamefont {Linhao}\
  \bibnamefont {Li}},\ }\bibfield  {title} {\enquote {\bibinfo {title}
  {{Duality viewpoint of noninvertible symmetry protected topological
  phases}},}\ }\href@noop {} {\  (\bibinfo {year} {2025}{\natexlab{a}})},\
  \Eprint {http://arxiv.org/abs/2502.20435} {arXiv:2502.20435
  [cond-mat.str-el]} \BibitemShut {NoStop}%
\bibitem [{\citenamefont {Shao}(2023)}]{Shao:2023gho}%
  \BibitemOpen
  \bibfield  {author} {\bibinfo {author} {\bibfnamefont {Shu-Heng}\
  \bibnamefont {Shao}},\ }\bibfield  {title} {\enquote {\bibinfo {title}
  {{What's Done Cannot Be Undone: TASI Lectures on Non-Invertible
  Symmetries}},}\ }\href@noop {} {\  (\bibinfo {year} {2023})},\ \Eprint
  {http://arxiv.org/abs/2308.00747} {arXiv:2308.00747 [hep-th]} \BibitemShut
  {NoStop}%
\bibitem [{\citenamefont {Kobayashi}\ \emph {et~al.}(2025)\citenamefont
  {Kobayashi}, \citenamefont {Nishioka}, \citenamefont {Otsuka},\ and\
  \citenamefont {Tanimoto}}]{Kobayashi:2025znw}%
  \BibitemOpen
  \bibfield  {author} {\bibinfo {author} {\bibfnamefont {Tatsuo}\ \bibnamefont
  {Kobayashi}}, \bibinfo {author} {\bibfnamefont {Yume}\ \bibnamefont
  {Nishioka}}, \bibinfo {author} {\bibfnamefont {Hajime}\ \bibnamefont
  {Otsuka}}, \ and\ \bibinfo {author} {\bibfnamefont {Morimitsu}\ \bibnamefont
  {Tanimoto}},\ }\bibfield  {title} {\enquote {\bibinfo {title} {{More about
  quark Yukawa textures from selection rules without group actions}},}\
  }\href@noop {} {\  (\bibinfo {year} {2025})},\ \Eprint
  {http://arxiv.org/abs/2503.09966} {arXiv:2503.09966 [hep-ph]} \BibitemShut
  {NoStop}%
\bibitem [{\citenamefont {Kobayashi}\ \emph {et~al.}(2024)\citenamefont
  {Kobayashi}, \citenamefont {Otsuka},\ and\ \citenamefont
  {Tanimoto}}]{Kobayashi:2024cvp}%
  \BibitemOpen
  \bibfield  {author} {\bibinfo {author} {\bibfnamefont {Tatsuo}\ \bibnamefont
  {Kobayashi}}, \bibinfo {author} {\bibfnamefont {Hajime}\ \bibnamefont
  {Otsuka}}, \ and\ \bibinfo {author} {\bibfnamefont {Morimitsu}\ \bibnamefont
  {Tanimoto}},\ }\bibfield  {title} {\enquote {\bibinfo {title} {{Yukawa
  textures from non-invertible symmetries}},}\ }\href {\doibase
  10.1007/JHEP12(2024)117} {\bibfield  {journal} {\bibinfo  {journal} {JHEP}\
  }\textbf {\bibinfo {volume} {12}},\ \bibinfo {pages} {117} (\bibinfo {year}
  {2024})},\ \Eprint {http://arxiv.org/abs/2409.05270} {arXiv:2409.05270
  [hep-ph]} \BibitemShut {NoStop}%
\bibitem [{\citenamefont {Choi}\ \emph
  {et~al.}(2022{\natexlab{b}})\citenamefont {Choi}, \citenamefont {Lam},\ and\
  \citenamefont {Shao}}]{Choi:2022jqy}%
  \BibitemOpen
  \bibfield  {author} {\bibinfo {author} {\bibfnamefont {Yichul}\ \bibnamefont
  {Choi}}, \bibinfo {author} {\bibfnamefont {Ho~Tat}\ \bibnamefont {Lam}}, \
  and\ \bibinfo {author} {\bibfnamefont {Shu-Heng}\ \bibnamefont {Shao}},\
  }\bibfield  {title} {\enquote {\bibinfo {title} {{Noninvertible Global
  Symmetries in the Standard Model}},}\ }\href {\doibase
  10.1103/PhysRevLett.129.161601} {\bibfield  {journal} {\bibinfo  {journal}
  {Phys. Rev. Lett.}\ }\textbf {\bibinfo {volume} {129}},\ \bibinfo {pages}
  {161601} (\bibinfo {year} {2022}{\natexlab{b}})},\ \Eprint
  {http://arxiv.org/abs/2205.05086} {arXiv:2205.05086 [hep-th]} \BibitemShut
  {NoStop}%
\bibitem [{\citenamefont {Cordova}\ and\ \citenamefont
  {Koren}(2023)}]{Cordova:2022qtz}%
  \BibitemOpen
  \bibfield  {author} {\bibinfo {author} {\bibfnamefont {Clay}\ \bibnamefont
  {Cordova}}\ and\ \bibinfo {author} {\bibfnamefont {Seth}\ \bibnamefont
  {Koren}},\ }\bibfield  {title} {\enquote {\bibinfo {title} {{Higher Flavor
  Symmetries in the Standard Model}},}\ }\href {\doibase
  10.1002/andp.202300031} {\bibfield  {journal} {\bibinfo  {journal} {Annalen
  Phys.}\ }\textbf {\bibinfo {volume} {535}},\ \bibinfo {pages} {2300031}
  (\bibinfo {year} {2023})},\ \Eprint {http://arxiv.org/abs/2212.13193}
  {arXiv:2212.13193 [hep-ph]} \BibitemShut {NoStop}%
\bibitem [{\citenamefont {Cordova}\ and\ \citenamefont
  {Ohmori}(2023)}]{Cordova:2022ieu}%
  \BibitemOpen
  \bibfield  {author} {\bibinfo {author} {\bibfnamefont {Clay}\ \bibnamefont
  {Cordova}}\ and\ \bibinfo {author} {\bibfnamefont {Kantaro}\ \bibnamefont
  {Ohmori}},\ }\bibfield  {title} {\enquote {\bibinfo {title} {{Noninvertible
  Chiral Symmetry and Exponential Hierarchies}},}\ }\href {\doibase
  10.1103/PhysRevX.13.011034} {\bibfield  {journal} {\bibinfo  {journal} {Phys.
  Rev. X}\ }\textbf {\bibinfo {volume} {13}},\ \bibinfo {pages} {011034}
  (\bibinfo {year} {2023})},\ \Eprint {http://arxiv.org/abs/2205.06243}
  {arXiv:2205.06243 [hep-th]} \BibitemShut {NoStop}%
\bibitem [{\citenamefont {Cordova}\ \emph
  {et~al.}(2024{\natexlab{a}})\citenamefont {Cordova}, \citenamefont {Hong},
  \citenamefont {Koren},\ and\ \citenamefont {Ohmori}}]{Cordova:2022fhg}%
  \BibitemOpen
  \bibfield  {author} {\bibinfo {author} {\bibfnamefont {Clay}\ \bibnamefont
  {Cordova}}, \bibinfo {author} {\bibfnamefont {Sungwoo}\ \bibnamefont {Hong}},
  \bibinfo {author} {\bibfnamefont {Seth}\ \bibnamefont {Koren}}, \ and\
  \bibinfo {author} {\bibfnamefont {Kantaro}\ \bibnamefont {Ohmori}},\
  }\bibfield  {title} {\enquote {\bibinfo {title} {{Neutrino Masses from
  Generalized Symmetry Breaking}},}\ }\href {\doibase
  10.1103/PhysRevX.14.031033} {\bibfield  {journal} {\bibinfo  {journal} {Phys.
  Rev. X}\ }\textbf {\bibinfo {volume} {14}},\ \bibinfo {pages} {031033}
  (\bibinfo {year} {2024}{\natexlab{a}})},\ \Eprint
  {http://arxiv.org/abs/2211.07639} {arXiv:2211.07639 [hep-ph]} \BibitemShut
  {NoStop}%
\bibitem [{\citenamefont {Cordova}\ \emph
  {et~al.}(2024{\natexlab{b}})\citenamefont {Cordova}, \citenamefont {Hong},\
  and\ \citenamefont {Koren}}]{Cordova:2024ypu}%
  \BibitemOpen
  \bibfield  {author} {\bibinfo {author} {\bibfnamefont {Clay}\ \bibnamefont
  {Cordova}}, \bibinfo {author} {\bibfnamefont {Sungwoo}\ \bibnamefont {Hong}},
  \ and\ \bibinfo {author} {\bibfnamefont {Seth}\ \bibnamefont {Koren}},\
  }\bibfield  {title} {\enquote {\bibinfo {title} {{Non-Invertible Peccei-Quinn
  Symmetry and the Massless Quark Solution to the Strong CP Problem}},}\
  }\href@noop {} {\  (\bibinfo {year} {2024}{\natexlab{b}})},\ \Eprint
  {http://arxiv.org/abs/2402.12453} {arXiv:2402.12453 [hep-ph]} \BibitemShut
  {NoStop}%
\bibitem [{\citenamefont {Choi}\ \emph {et~al.}(2024)\citenamefont {Choi},
  \citenamefont {Forslund}, \citenamefont {Lam},\ and\ \citenamefont
  {Shao}}]{Choi:2023pdp}%
  \BibitemOpen
  \bibfield  {author} {\bibinfo {author} {\bibfnamefont {Yichul}\ \bibnamefont
  {Choi}}, \bibinfo {author} {\bibfnamefont {Matthew}\ \bibnamefont
  {Forslund}}, \bibinfo {author} {\bibfnamefont {Ho~Tat}\ \bibnamefont {Lam}},
  \ and\ \bibinfo {author} {\bibfnamefont {Shu-Heng}\ \bibnamefont {Shao}},\
  }\bibfield  {title} {\enquote {\bibinfo {title} {{Quantization of Axion-Gauge
  Couplings and Noninvertible Higher Symmetries}},}\ }\href {\doibase
  10.1103/PhysRevLett.132.121601} {\bibfield  {journal} {\bibinfo  {journal}
  {Phys. Rev. Lett.}\ }\textbf {\bibinfo {volume} {132}},\ \bibinfo {pages}
  {121601} (\bibinfo {year} {2024})},\ \Eprint
  {http://arxiv.org/abs/2309.03937} {arXiv:2309.03937 [hep-ph]} \BibitemShut
  {NoStop}%
\bibitem [{\citenamefont {Kobayashi}\ and\ \citenamefont
  {Otsuka}(2024)}]{Kobayashi:2024yqq}%
  \BibitemOpen
  \bibfield  {author} {\bibinfo {author} {\bibfnamefont {Tatsuo}\ \bibnamefont
  {Kobayashi}}\ and\ \bibinfo {author} {\bibfnamefont {Hajime}\ \bibnamefont
  {Otsuka}},\ }\bibfield  {title} {\enquote {\bibinfo {title} {{Non-invertible
  flavor symmetries in magnetized extra dimensions}},}\ }\href {\doibase
  10.1007/JHEP11(2024)120} {\bibfield  {journal} {\bibinfo  {journal} {JHEP}\
  }\textbf {\bibinfo {volume} {11}},\ \bibinfo {pages} {120} (\bibinfo {year}
  {2024})},\ \Eprint {http://arxiv.org/abs/2408.13984} {arXiv:2408.13984
  [hep-th]} \BibitemShut {NoStop}%
\bibitem [{\citenamefont {Cao}\ \emph {et~al.}(2025{\natexlab{b}})\citenamefont
  {Cao}, \citenamefont {Miao},\ and\ \citenamefont {Yamazaki}}]{Cao:2025qnc}%
  \BibitemOpen
  \bibfield  {author} {\bibinfo {author} {\bibfnamefont {Weiguang}\
  \bibnamefont {Cao}}, \bibinfo {author} {\bibfnamefont {Yuan}\ \bibnamefont
  {Miao}}, \ and\ \bibinfo {author} {\bibfnamefont {Masahito}\ \bibnamefont
  {Yamazaki}},\ }\bibfield  {title} {\enquote {\bibinfo {title} {{Global
  symmetries of quantum lattice models under non-invertible dualities}},}\
  }\href@noop {} {\  (\bibinfo {year} {2025}{\natexlab{b}})},\ \Eprint
  {http://arxiv.org/abs/2501.12514} {arXiv:2501.12514 [cond-mat.str-el]}
  \BibitemShut {NoStop}%
\bibitem [{\citenamefont {'t~Hooft}(1980)}]{tHooft:1979rat}%
  \BibitemOpen
  \bibfield  {author} {\bibinfo {author} {\bibfnamefont {Gerard}\ \bibnamefont
  {'t~Hooft}},\ }\bibfield  {title} {\enquote {\bibinfo {title} {{Naturalness,
  chiral symmetry, and spontaneous chiral symmetry breaking}},}\ }\href
  {\doibase 10.1007/978-1-4684-7571-5_9} {\bibfield  {journal} {\bibinfo
  {journal} {NATO Sci. Ser. B}\ }\textbf {\bibinfo {volume} {59}},\ \bibinfo
  {pages} {135--157} (\bibinfo {year} {1980})}\BibitemShut {NoStop}%
\bibitem [{\citenamefont {Tanimoto}\ and\ \citenamefont
  {Yanagida}(2016)}]{Tanimoto:2016rqy}%
  \BibitemOpen
  \bibfield  {author} {\bibinfo {author} {\bibfnamefont {Morimitsu}\
  \bibnamefont {Tanimoto}}\ and\ \bibinfo {author} {\bibfnamefont {Tsutomu~T.}\
  \bibnamefont {Yanagida}},\ }\bibfield  {title} {\enquote {\bibinfo {title}
  {{Occam's Razor in Quark Mass Matrices}},}\ }\href {\doibase
  10.1093/ptep/ptw024} {\bibfield  {journal} {\bibinfo  {journal} {PTEP}\
  }\textbf {\bibinfo {volume} {2016}},\ \bibinfo {pages} {043B03} (\bibinfo
  {year} {2016})},\ \Eprint {http://arxiv.org/abs/1601.04459} {arXiv:1601.04459
  [hep-ph]} \BibitemShut {NoStop}%
\bibitem [{\citenamefont {Hiller}\ and\ \citenamefont
  {Schmaltz}(2001)}]{Hiller:2001qg}%
  \BibitemOpen
  \bibfield  {author} {\bibinfo {author} {\bibfnamefont {Gudrun}\ \bibnamefont
  {Hiller}}\ and\ \bibinfo {author} {\bibfnamefont {Martin}\ \bibnamefont
  {Schmaltz}},\ }\bibfield  {title} {\enquote {\bibinfo {title} {{Solving the
  Strong CP Problem with Supersymmetry}},}\ }\href {\doibase
  10.1016/S0370-2693(01)00814-0} {\bibfield  {journal} {\bibinfo  {journal}
  {Phys. Lett. B}\ }\textbf {\bibinfo {volume} {514}},\ \bibinfo {pages}
  {263--268} (\bibinfo {year} {2001})},\ \Eprint
  {http://arxiv.org/abs/hep-ph/0105254} {arXiv:hep-ph/0105254} \BibitemShut
  {NoStop}%
\bibitem [{\citenamefont {Tanimoto}\ and\ \citenamefont
  {Yanagida}(2024)}]{Tanimoto:2024nwm}%
  \BibitemOpen
  \bibfield  {author} {\bibinfo {author} {\bibfnamefont {Morimitsu}\
  \bibnamefont {Tanimoto}}\ and\ \bibinfo {author} {\bibfnamefont {Tsutomu~T.}\
  \bibnamefont {Yanagida}},\ }\bibfield  {title} {\enquote {\bibinfo {title}
  {{Prediction of the CP Phase $\delta_{CP}$ in the Neutrino Oscillation and an
  Axion-less Solution to the Strong CP Problem}},}\ }\href@noop {} {\
  (\bibinfo {year} {2024})},\ \Eprint {http://arxiv.org/abs/2410.01224}
  {arXiv:2410.01224 [hep-ph]} \BibitemShut {NoStop}%
\bibitem [{\citenamefont {Tanimoto}\ and\ \citenamefont
  {Yanagida}(2025)}]{Tanimoto:2025fnj}%
  \BibitemOpen
  \bibfield  {author} {\bibinfo {author} {\bibfnamefont {Morimitsu}\
  \bibnamefont {Tanimoto}}\ and\ \bibinfo {author} {\bibfnamefont {Tsutomu~T.}\
  \bibnamefont {Yanagida}},\ }\bibfield  {title} {\enquote {\bibinfo {title}
  {{Axionless Solution to the Strong CP Problem -- two-zeros textures of the
  quark and lepton mass matrices and neutrino CP violation --}},}\ }\href@noop
  {} {\  (\bibinfo {year} {2025})},\ \Eprint {http://arxiv.org/abs/2504.06599}
  {arXiv:2504.06599 [hep-ph]} \BibitemShut {NoStop}%
\bibitem [{\citenamefont {Branco}\ \emph {et~al.}(2012)\citenamefont {Branco},
  \citenamefont {Ferreira}, \citenamefont {Lavoura}, \citenamefont {Rebelo},
  \citenamefont {Sher},\ and\ \citenamefont {Silva}}]{Branco:2011iw}%
  \BibitemOpen
  \bibfield  {author} {\bibinfo {author} {\bibfnamefont {G.~C.}\ \bibnamefont
  {Branco}}, \bibinfo {author} {\bibfnamefont {P.~M.}\ \bibnamefont
  {Ferreira}}, \bibinfo {author} {\bibfnamefont {L.}~\bibnamefont {Lavoura}},
  \bibinfo {author} {\bibfnamefont {M.~N.}\ \bibnamefont {Rebelo}}, \bibinfo
  {author} {\bibfnamefont {Marc}\ \bibnamefont {Sher}}, \ and\ \bibinfo
  {author} {\bibfnamefont {Joao~P.}\ \bibnamefont {Silva}},\ }\bibfield
  {title} {\enquote {\bibinfo {title} {{Theory and phenomenology of
  two-Higgs-doublet models}},}\ }\href {\doibase 10.1016/j.physrep.2012.02.002}
  {\bibfield  {journal} {\bibinfo  {journal} {Phys. Rept.}\ }\textbf {\bibinfo
  {volume} {516}},\ \bibinfo {pages} {1--102} (\bibinfo {year} {2012})},\
  \Eprint {http://arxiv.org/abs/1106.0034} {arXiv:1106.0034 [hep-ph]}
  \BibitemShut {NoStop}%
\end{thebibliography}%

\end{document}